# Thin Film growth of Solid State materials

A.S Bhattacharyya[1,2*], P. Prabhakar[1], R.P. Kumar[1], S. Sharma[3], S.K.Raj[1], R.Ratn[1], P.Kommu[1]

[1]Centre for Nanotechnology; [2]Centre of Excellence in Green and Efficient Energy Technology (CoE-GEET) and [3]Centre for Water Engineering and Management,
Central University of Jharkhand, Ranchi – 835 205, India

Corresponding author Email: 2006asb@gmail.com ; arnab.bhattacharya@cuj.ac.in

*Abstract*

*Magnetron sputtering has also been used to deposit thin films of some materials and it has significant technological importance. A modeling on deposition of epitaxial thin films of Yttrium Stabilized Zirconia (YSZ) was done the diffusion of adatom on the surface were studies. There exists a strong interaction of ions formed in the plasma during the sputtering process. Cu thin films were deposited on Si. Nanocomposite thin film of SiCN showed dendritic growth.*

**Keywords**
Yttrium Stabilized Zirconia; epitaxial; sputtering; Cu; SiCN

**Introduction**

Magnetron sputtering is an effective means of depositing nanocomposite and epitaxial thin films. An attempt was made to computationally simulate the sputtering process and deposit these films and study the diffusion of the adatom on the surface [1]. Epitaxial thin films of materials used in solid state ionics like Yttrium Stabilized Zirconia (YSZ), Rare Earth Oxides (REO) are suitable systems to study the interfacial ion transport. YSZ is an efficient oxygen ion conductor, vacancy present are responsible for ionic transport. By altering the structure we can increase the ionic conductivity [2].

There exists a strong interaction between the ions formed in the plasma during the sputtering process, the adatoms and the sputtering parameters which also affects the performance of the device, which can be batteries, solid oxide fuel cell (SOFC), ionic membranes, sensors or ceramic oxygen generators (COG). The introduction of gases for reactive sputtering also complicates the system. Copper in the form of thin film has major applications in the electronic industry for interconnections and device fabrication as well as in the medical field [3-5].

Si-C-N ceramic coatings of ternary nanocomposites have shown newer and improved mechanical and functional properties over the coarser and monolithic coatings. Properties such as high hardness, wear resistance, oxidation resistance, tuneable band gap and chemical inertness have been observed for Si-C-N which makes it potentially useful for numerous applications. A way of mixing the three atomic components viz. Si, C and N is by magnetron sputtering i.e using SiC targets in argon/nitrogen plasma. This method allows one to produce nanocrystallites of SiC, $Si_3N_4$, $C_3N_4$ as well as nanocrystalline graphite on Si–N–C amorphous matrix. Correlations of structural and mechanical properties of SiCN nanocomposite coatings deposited by magnetron sputtering have been extensively done and published [6-12].

**Materials and Methods**

The rate of sputtering i.e. the amount of target materials sputtered per unit time is given in Equation 1, where M is the molar weight of the target [kg/mol]; r is the density of the material [kg/m$^3$]; NA is the Avogadro number, $6.02 \times 10^{26}$ 1/kmol; e is the electron charge, $1.6 \times 10^{-19}$ As ; S is the sputtering yield (atom/ion) and $j_P$ is the primary ion current density [A/m$^2$]. This rate (z/t) divided by the atomic diameter which gives us the rate of sputtering in terms of atomic layers (AL) per second [13].

$$z/t = M/(rNAe)\ S.\ j_P \quad (1)$$

The adatoms sputtered from the target are multiplied with exp (-d/L) to get rough estimates of the deposited atomic layers per second. L (cm) is the mean free path of the sputtered atom, which is related to the sputtering pressure P (Torr) and the molecular diameter Da of the reactive gas (252 pm for $O_2$)

*$L = 2.303 \times 10^{-20} T/(PDa^2)$*

Where T is the deposition temperature and d (cm) is the distance traversed by the adatom for deposition which was taken as 30 cm considering the target the substrate distance [14].

The rate of deposition of $ZrO_2$ at 600 eV Ar ion bombardment is 40 Å/s. Yttrium on the other hand has a sputtering rate of 85 Å/s under the same condition [15]. For the YSZ thin film we will consider the slower rate of deposition i.e. 40 Å/s. The atomic radii of Zr is 206 pm and that of Y is 212 pm. As the film is mainly $ZrO_2$ with Y as additive, we will consider 206 pm for atomic layer calculation. The atomic layers were calculated from 1 s to 100 s. The deposition rate shows abrupt changes with oxygen flow rates [16].

We must understand that the rate of sputtering and rate of deposition are two different parameters. The adatoms from the target will be acted upon by oxygen and finally deposited. The oxide formation therefore usually takes place during the traversing of adatoms from the source to the target.

Going backwards as per our proposed model, we can estimate the rate of sputtering from the rate of deposition, which comes out to be the same as the rate of deposition. Thus the change in deposition rate is due to oxygen flow and its reaction with the adatoms after they have been ejected out from the target.

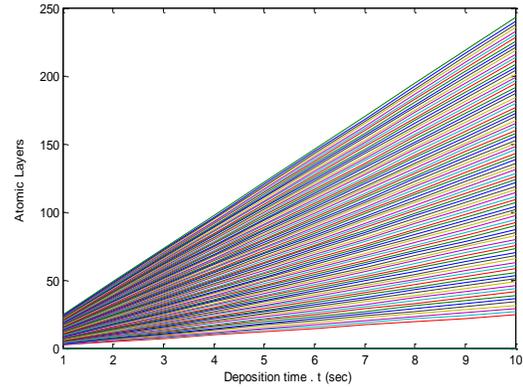

**Figure 1**: Atomic layers of $ZrO_2$ deposited with rate of sputtering varying from 10 to 100 Å/s

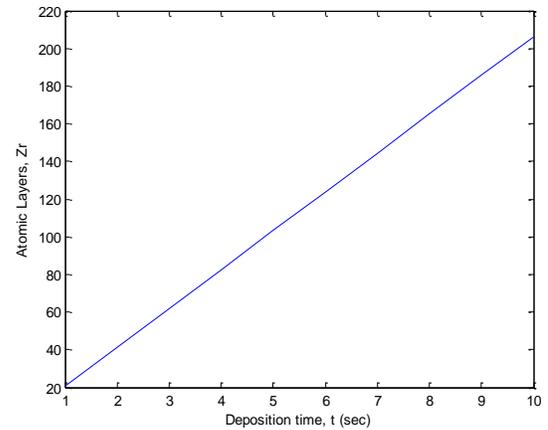

**Figure 2**. Linear variation of the atomic Layers of Zr with deposition time

The sputtering rate of the compound material formed by reaction of oxygen introduced in the sputtering chamber with the target is much less than that of target itself. The target is also called poisoned target which has been dealt by us previously while modeling sputter deposition of TiOx films [17]. A change in phase is reported due to oxygen flow rate difference during sputtering where also a sharp transition in deposition rate is observed [18].

## Results

The variation of atomic layers of ZrO$_2$ deposited with the rate of sputtering varying from 10 to 100 Å/s is shown in Figure 1. As proposed by Bhattacharyya [6], the rate is almost halved suddenly when there is an oxide formation. We assume a change of rate from 80 Å/s to 40 Å/s. The sputter rate of element Zr is 85 Å/s which confirms this fact. A linear variation of the atomic layers of Zr with deposition time is shown in Figure 2.

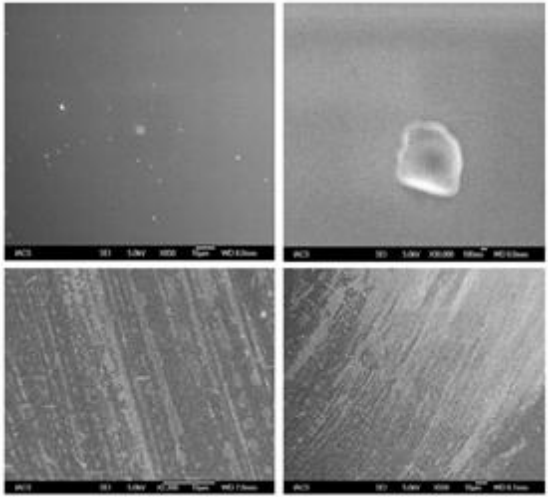

**Figure 3** : SEM of Cu deposited on Si

Copper deposited on Si(100) and Si(111) has been studied and published by our group [2, 21]. An SEM (scanning electron microscopy) of Cu on Si is shown in Figure 3. A change in morphology is observed in the deposited film with varying deposition conditions. The growth has been reported to be of Volmer-Weber (3D island) type. The islands seem to form a series of arrays which in this case we call nanowires (NW).

Cu NW on silicon substrates is useful for interconnects in electronic devices [22]. The Cu NW/ Si system has also been found to be causing an enhancement in pool boiling heat transfer compared to bare substrate and is applicable in small scale electronic devices [23].

A rare dendritic growth has been observed in some areas in SiCN and CN films deposited by RF magnetron sputtering and plasma enhanced CVD respectively on Si (100) substrates (Figure 2). The rapid rate of nucleation and growth led to instabilities in the growth pattern and the surface energy release rate was more through convection than diffusion It opens up new fields of fractal study in the case of CN and SiCN based materials and thin films. Dendritic growth usually occurs in metals due to interfacial interaction between the molten state and the solid state.

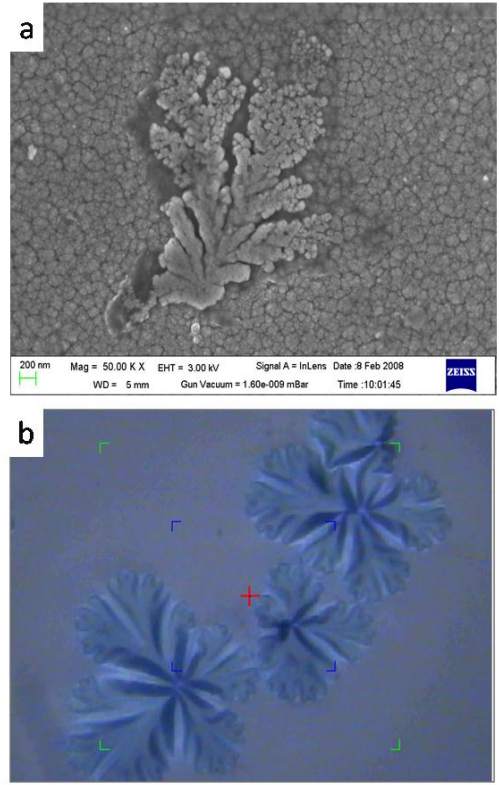

**Figure 4**: Dendritic growth observed in (a) PVD grown SiCN films and (b) CVD grown CN films

Fractal growth in carbon-based organic thin films is quite uncommon and not well understood. Carbon nitride and silicon carbon nitride based materials and films have shown novel multifunctional properties and have been prepared through various techniques. However, to the best of our knowledge, no reports showing dendrite growth in SiCN or CN films are present till date. A SiCN thin film, although not truly an organic film, has a resemblance in properties to hydrocarbon thin films as a small fraction of hydrogen is present in the top surface layer due to trapped residual water molecules.

These hydrogen atoms can react with carbon present in the film. The presence of nitrogen below 20 at % in the case of CNx films on the other hand, makes it equivalent to a nitrogen free carbon film. The formation of a CH compound is therefore quite possible at the surface of the SiCN film. Recent reports have shown fractal growth of hydrocarbon pentacene ($C_{14}H_{22}$) on Si (100) substrates which finds application in organic thin film devices [19].

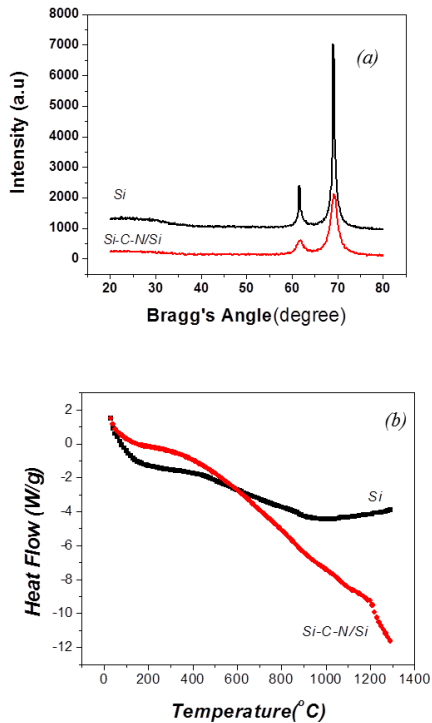

**Figure 5**: (a) XRD and (b) TGA DSC study of SiCN/Si and Si substrate

Dendritic growths were observed in SiCN films grown by magnetron sputtering at high substrate temperature of 400°C. The micrographs were captures via FESEM (Zeiss, Germany) as given in Figure 4.The rapid rate of nucleation and growth process led to instabilities in the growth pattern and the surface energy release rate was more through convection than diffusion. The nature of dendritic growth was however different for CVD grown CNx films. The dendrites were much larger in size and were associated with film delamination form the substrate most probably due to high amount of stress being generated in the film. On close observation of the dendritic SiCN structure one can see that there are two different regimes. The first one starts from the bottom to the middle where the nanograins are not distinctly visible. In the other regime the grains are prominent which starts in the middle and extends upto the branched structure.

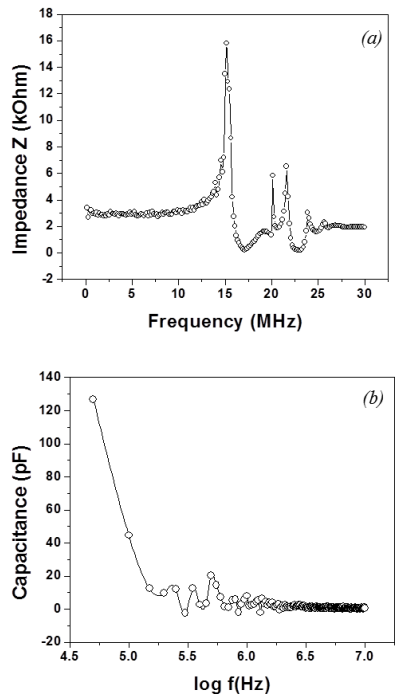

**Figure 6**: Impedance analysis of SiCN thin films

The dielectric properties are measured using Hioki 3532 LCR Hightester, which is an impedance meter (Fig 6). Impedance analysis of similar polymer derived silicon oxycarbides have been done previously and the increase in conductivity of both matrix and free carbon phase was found to correlate with the increase with in temperature [23]. This section is stored in vixra.org [13]

## Discussion

The difference between the atomic layers of Zr and $ZrO_2$ is the effect of oxygen flow. The variation of differences in atomic layers with deposition time and its derivative giving the oxygen deposition rate is shown in Figure 7. Interestingly, instead of one abrupt change as reported before, we observed more than one abrupt change. Currently, we can say that the sudden decrease corresponds to oxide adsorption and desorption of oxygen on the surface. The variation has an oscillatory nature as shown in Figure 8.

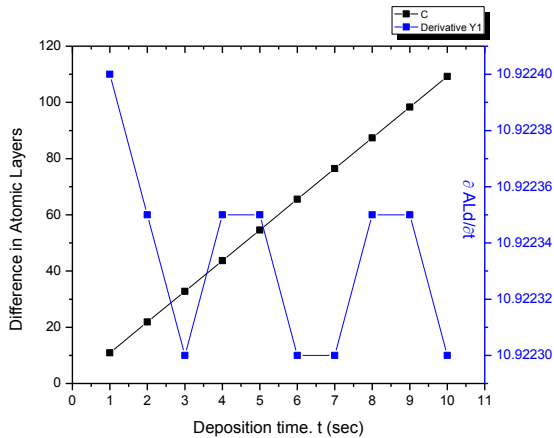

**Figure 7**: Variation of atomic layers difference with deposition time and its derivative

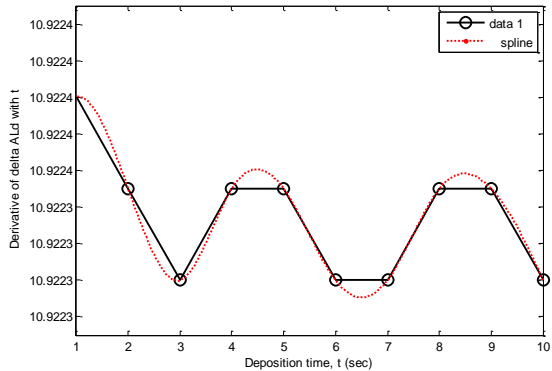

Fig 8: Oscillatory nature of the variation of derivative of atomic layers difference w.r.t deposition time

In the SiCN films, the reason behind having two different structural natures can be attributed to the diffusion of adatoms during film growth. Dendritic growth usually takes place during solidification of a supercooled liquid. Supercooling is a process of lowering the temperature of a liquid below its melting point without solidification. As the liquid solidifies, the dendrite grows in structure. Perhaps, a large temperature gradient might have risen during the thin film growth which melts a part of the film. The melted part, going through the process of solidification again, could have given the dendritic structure. The parameter which defines non-dimensional super cooling is called the Stefan number $S_t$ [20]. There are reports of dendritic growth in thin films where temperature and film thickness have been found to contribute significantly [21]. XRD and TGA DSC studies of the films are given in figure 5. Previous studies have shown a 4-stage weight loss on polymer derived SiCN and it has been found to be thermally stable at high temperatures [22].

## Conclusion

The results show that there exists a strong interaction between the ions formed in the plasma during the sputtering process and the adatoms as well as the reactive gases introduced. The phenomenon of adsorption and surface diffusion in this case needs to be looked into. The islands seem to form a series of arrays in Cu/Si sputtered systems which might be helpful in electronic devices. The dendritic growth in SiCN films might have a greater impact in this context, being an optoelectronic material itself with a tuneable band gap. It opens up a new field of fractal studies in the case of CN and SiCN based materials and thin films.

The article is also self-archived references [19] and [20]
## References

1. AS Bhattacharyya, S Kumar, S Jana, PY Kommu, K Gaurav, S Prabha, VS.Kujur, P.Bharadwaj, Sputter based Epitaxial Growth and Modeling of Cu/Si Thin Films, *Int. J. Thin. Fil. Sci. Tec* 4 (3), 2015, 173-177.
2. M. T. Elm, J. D. Hofmann, C. Suchomski, J. Janek, T.Brezesinski, *ACS Appl. Mater. Interfaces* 2015, 7, 11792−11801 and references within.
5